\def\zz{\begin{equation}}
\def\prob{{\cal P}}
\def\sign{{\rm sign}}

\documentstyle{ioplppt}
\jl{1} 

\begin{document}
\input epsf


\title{Stationary definition of persistence for finite temperature phase
ordering }

\author{J-M~Drouffe\dag\, and C~Godr\`eche\ddag \S}

\address{\dag\ Service de Physique Th\'eorique,
CEA-Saclay, 91191 Gif sur Yvette cedex, France}

\address{\ddag\ Service de Physique de l'\'Etat Condens\'e,
CEA-Saclay, 91191 Gif sur Yvette cedex, France}

\address{\S\ Laboratoire de Physique Th\'eorique et Mod\'elisation,
Universit\'e de Cergy Pontoise, France}

\begin{abstract}
For the two dimensional kinetic Ising model at finite temperature, the
local mean magnetisation
$M_t=t^{-1}\int_{0}^t\sigma(t')\d t'$, simply related to the fraction of time spent by
a given spin in the positive direction, has a limiting distribution, singular at
$\pm m_0(T)$, the Onsager spontaneous magnetization. 
The exponent of this singularity defines the persistence exponent $\theta$.
We also study first passage exponents associated to persistent large deviations of
$M_t$, and their temperature dependence.

\end{abstract}

{\it submitted for publication to \JPA}

\medskip

In this work we present a new approach to the study of persistence for
systems undergoing phase ordering \cite {langer,brayRevue} at finite temperature, which
we shall illustrate on the case of the two dimensional Ising model.
In this approach persistence appears as a stationary property of the coarsening system,
and the role of the Onsager spontaneous magnetization at equilibrium
$m_0(T)$ is made apparent, thus revealing new fundamental features of phase ordering.
It departs from previous approaches to finite temperature
persistence
\cite{derT,sirT,stauT,hin}, where these features did not appear. 

Consider a system of Ising spins $\sigma_i(t)=\pm1$ located at sites $i=1,\ldots,N$,
started from a random initial condition, and evolving
under the heat bath dynamics at fixed temperature below the critical temperature. 
At each time step a spin is picked at random, and
updated with the probability
\begin{equation}
P(\sigma_i(t+\d t)=+1)=\frac{1}{2}\left(1+\tanh \frac{1}{T}\sum_j\, \sigma_j(t)\right)
\label{eqheat}
\end{equation}
where the sum runs over the neighbours of site $i$.
Under this dynamics spins thermalize in their local
environment. 
Therefore the system coarsens, i.e. domains of opposite signs grow and, in the scaling
regime, the system is statistically self similar, with only one single characteristic
length scale, which is the size of a typical domain \cite{langer,brayRevue}. 

The question of persistence is to determine 
the fraction of space $R(t)$ which remained in the same phase up to time $t$
\cite{marcos,bray1} (or from time $t_1$ to time $t_2$). 
For the two dimensional Ising model at zero temperature, two phases coexist,
corresponding to all spins equal to $+1$ or all spins equal to $-1$.
Hence $R(t)$ is equivalently defined as the fraction of spins which did not flip up
to time $t$ \cite{der1}, i.e. which were not swept by an interface between domains of
all spins $+1$ or all spins $-1$.
Numerical measurements indicate an algebraic decay $R(t)\sim t^{-\theta}$, with
$\theta\approx 0.22$ \cite{der1,stau1,der2}. 

The definition of persistence at finite temperature below $T_c$ is more subtle to
implement because one has to make clear what is meant by `phase'.
By essence, in the coarsening process there is phase separation, each phase wanting to
develop at the expense of the other.
It is therefore intuitive that the system, though perpetually out of equilibrium, tries
to reach locally one of the two equilibrium phases, corresponding to $\pm m_0(T)$,
where
\begin{equation}
m_0(T)=\left(1-\left(\sinh \frac{2}{T}\right)^{-4}\right)^{1/8}
\label{eqons}
\end{equation}
is the Onsager spontaneous magnetization at equilibrium \cite{onsager}.
Hence in the scaling regime the average magnetization
inside a domain measured on a scale of time small compared to the flipping time of the
domain, should be close to the equilibrium magnetization at this temperature.
Coming back to persistence, the definition of $R(t)$ should reflect, in one
way or the other, the fact that a given point in space remained in a phase
of average magnetization equal to $\pm m_0(T)$ up to time $t$.
This intuitive analysis is confirmed by what follows.

The central point of our approach is to consider the statistics of the local mean
magnetization --simply related to the fraction of time spent by a spin in the positive
direction-- in the limit of large times.
This line of thought was already used in \cite{dg} in the study of domain coarsening for the
one dimensional Ising model at zero temperature, or for the simple diffusion equation
evolved from a random initial condition (see also \cite{new}). 
The idea is that since persistence probes the past history of the system,
a natural quantity to consider is 
$\int_0^t \sigma(t')\d t'=T_t^{+}-T_t^{-}$,
where $\sigma(t)$ is the spin at site $i$ and $T_t^{+}$ ($T_t^{-}$) is the length of
time spent by the spin pointing upward (downward), with $t=T_t^{+}+T_t^{-}$.
The local mean magnetization is defined as
\begin{equation}
M_t=\frac{1}{t}\int_0^t \sigma(t') \d t' =2 \frac{T_t^{+}}{t}-1
.\label{eqmag}
\end{equation}
For instance at zero temperature, the persistence probability $R(t)$ is equal
to $\prob(M_t=1)$ since the event $\{\sigma(t')=1, \forall t'\le t\}$ is
identical to the event $\{M_t=1\}$. 
Since $M_t$ is a local quantity varying from site to
site, one is naturally led to investigate the distribution of $M_t$,
\begin{equation}
P(t,x)=\prob(M_t\ge x)\qquad (-1\le x\le1)
.\label{eqptx}
\end{equation}
For the one dimensional Ising model at $T=0$ is was shown in \cite{dg} by
analytical arguments and numerical measurements that, when $t\to\infty$, $P(t,x)$
converges to a limit distribution $P_\infty(x)$ with density 
\begin{equation}
f_M(x)=-\frac{\d}{\d x}P_\infty(x)
\label{eqfm}
\end{equation}
singular at $x=\pm1$ with singularity exponent equal to $\theta-1$.
It is for instance easy to show that, when
$t\to\infty$, the limit of $\langle M_t^2\rangle$ is a constant equal to $\hat A(1)$,
the Laplace transform of the two-time correlation with respect to the variable $\ln
t$, at argument equal to 1
\cite{dg}. 
This result therefore provides a {\it stationary} definition of
persistence at zero temperature. 
The same holds for the diffusion equation \cite{dg,new}.

\medskip
We now address the same questions at finite temperature.
We first report on numerical results.
\begin{figure}
\begin{center}
\leavevmode
\epsfxsize=\hsize
\epsfbox{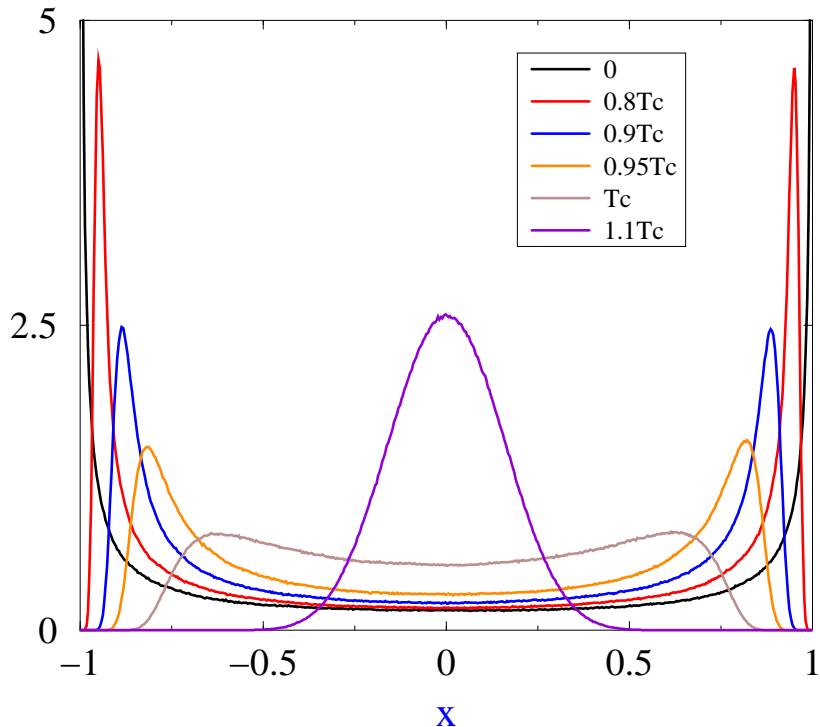}
\end{center}
\caption{
Histogram of the density of $M_t$ at time $t=1000$, for $T=0,\,
0.8 T_c, \, 0.9 T_c,\,0.95 T_c, \,T_c, \,1.1 T_c$, from bottom to top (at the central point $x=0$).
(The system size is $3072^2$.)}
\label{fig1}
\end{figure}
Figure 1 depicts the histogram of the density of $M_t$ at time $t=1000$ for values of
$T$ ranging from $0$ to $1.1 T_c$. 
Already for such a short time, and for every temperature $T<T_c$, the density is
maximum around
$\pm m_0(T)$, the equilibrium magnetization (\ref{eqons}). 
At larger times and for $T>T_c$, the density
of
$M_t$ becomes peaked around zero, i.e. the mean magnetization converges toward the
average magnetization per spin
$\langle\sigma\rangle=0$, reflecting the fact that the system reaches equilibrium.
At $T_c$ the peaking of the density of $M_t$ is observed to be very slow. 
Finally at $T<T_c$,  $P(t,x)$ converges,  when
$t\to\infty$, to a limit  distribution $P_\infty(x)$ with density $f_M(x)$ given
as in (\ref{eqfm}).

The existence of a limit law at finite temperature $T<T_C$ relies on the same
arguments as for the zero temperature case.
For instance the convergence of $\langle
M_t^2\rangle$ to a constant equal to $\hat A(1)$ still holds since it only
relies on the existence of a scaling regime \cite{dg}.
\begin{figure}
\begin{center}
\leavevmode
\epsfxsize=\hsize
\epsfbox{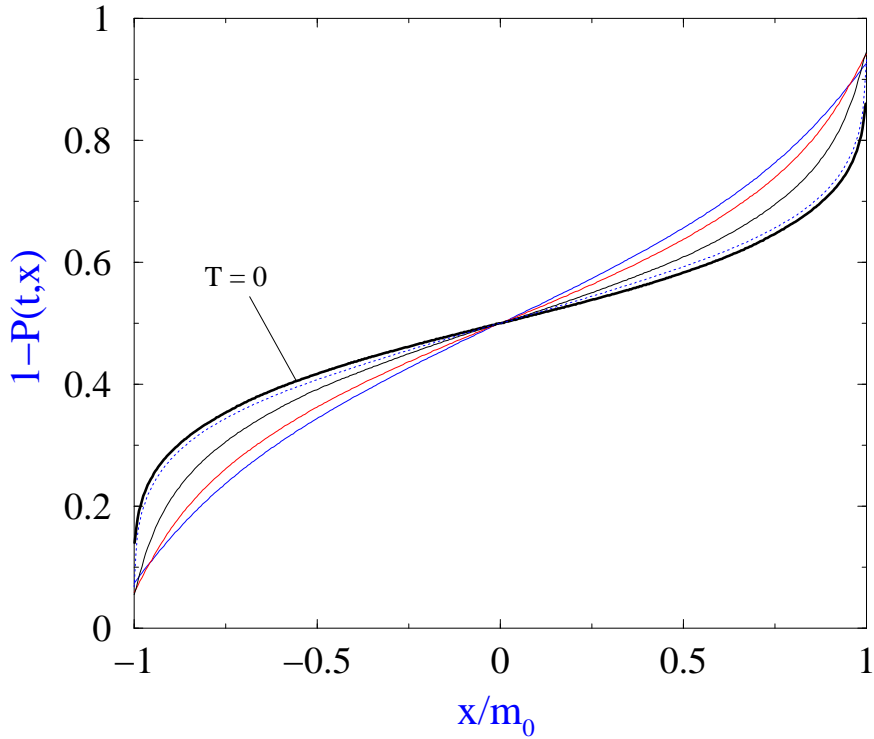}
\end{center}
\caption{
Plot of $1-P(t,x)$  against the rescaled variable $x/m_0(T)$, at  $T=0$ (full), $T=0.8\,T_c$ (dots), for $t=20\,000$,
and at $T=0.98\, T_c$, for $t=1000, 5000$ and $30\,000$.
(The system size is $3072^2$.)
\label{fig2}}
\end{figure}
The striking fact is that now the density concentrates on $[-m_0(T),m_0(T)]$
 with an exponential decay with time of
$\prob(M_t>m_0(T))$ to $0$. 
Moreover the limit density is singular at $\pm m_0(T)$, with a
singularity exponent $\theta-1$ which defines the persistence exponent at
temperature
$T$. 
 
This leads to the question of the temperature dependence of persistence for
$T<T_c$.
The simplest assumption is that, during the coarsening process, the scales of
time between a short time regime and the scaling regime decouple, yielding the
following relation between the moments of the limit distributions at $T$ and at zero
temperature,
\begin{equation}
\langle M^{2k}\rangle_T=(m_0(T))^{2k}\langle M^{2k}\rangle_0
\label{eqm2T}
\end{equation}
implying the identity of the limit
distributions, if $M_t$ is rescaled by
$m_0(T)$, and as
a consequence, the temperature independence of $\theta$.
This would be in agreement with
the usual view that zero temperature is an attracting fixed point for the dynamics of
phase ordering
\cite{brayRevue,sirT}. 
Equation (\ref{eqm2T}) is hard to check by numerical measurements because the convergence of 
the data is observed to be slow.
The difficulty is illustrated by figure 2 which
depicts a plot of $1-P(t,x)$  against the rescaled variable $x/m_0(T)$, at  $T=0$ and $0.8\,T_c$, for $t=20\,000$,
and at $T=0.98\, T_c$, for $t=1000, 5000$ and $30\,000$.
Though one cannot be conclusive on the sole basis of numerical measurements, data collapse seems nevertheless plausible.
Let us note that the limit distribution $1-P_\infty(x)$ at $T=0$ is well
approximated by a beta distribution, as was observed for the $1d$ Ising model \cite{dg},
or for the diffusion equation \cite{dg,new}.
The singularity exponent of the beta distribution is found to be around $0.22$.
A more precise numerical determination of the exponent
$\theta$ from the limit distribution of
$M_t$ needs further work and will be presented elsewhere. 

Let us summarize at this point. 
For $T<T_c$, the local mean magnetization
$M_t$  has a limit probability density when
$t\to\infty$, defined on the interval $[-m_0(T),m_0(T)]$, where $m_0(T)$ is the
equilibrium magnetization, and singular at both ends.
This provides a {\it stationary}
definition of persistence, which is a natural extension of the zero temperature case,
where the singularity exponent defines the persistence exponent. 
These are the
central results of the present work. 

\begin{figure}
\begin{center}
\leavevmode
\epsfxsize=\hsize
\epsfbox{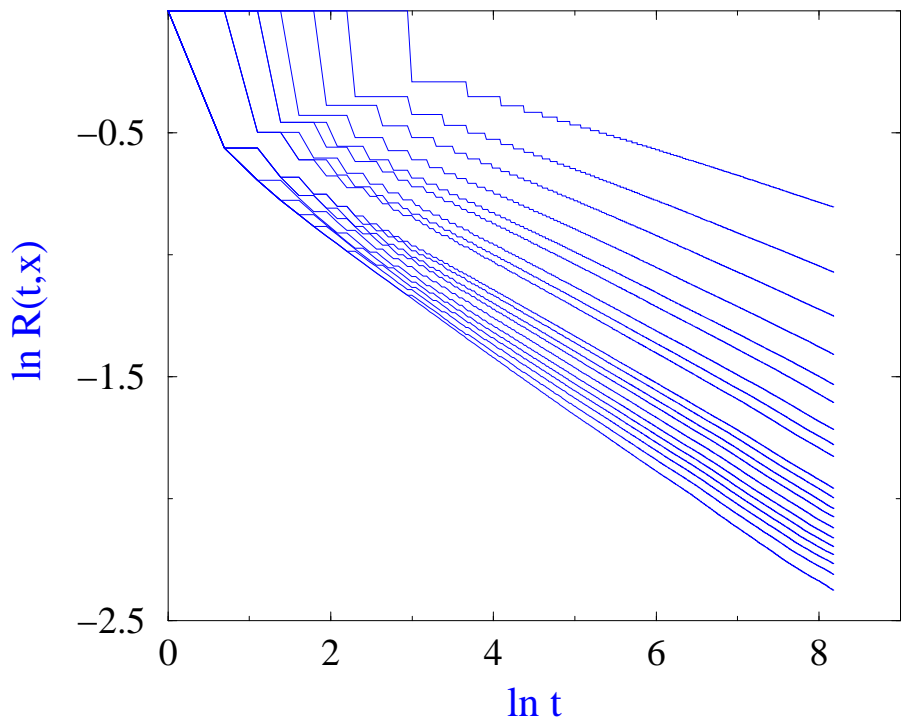}
\end{center}
\caption{
Log-log plot of $R(t,x)$ for the two dimensional Ising model at zero temperature, $x$
varying from $-1$ to $1$, by steps of $0.1$, from top to bottom. 
(The system size is $1536^2$.)
\label{fig3}}
\end{figure}

\medskip
Another new aspect of persistence introduced in \cite{dg} is concerned with
{\it persistent large deviations}.
The probability of persistent large deviations above the level $x$ ($-1\le
x\le 1$), denoted by $R(t,x)$, is defined as the probability that the
mean magnetization was, for all previous times, greater than $x$ \cite{dg},
\begin{equation}
R(t,x)=\prob(M_{t'}\ge x, \forall t'\le t)
.\end{equation}
In other words one is interested in the persistence probability of the stochastic
process $\sigma(t,x)=\sign(M_t-x)$ \cite{dg}.
If one views the stochastic process $\sigma(t)$ as the successive steps of a
fictitious random walker, then
$M_t$ is the mean speed of the walker between 0 and $t$, and $R(t,x)$ is the
probability that this mean speed remained larger than $x$ between 0 and $t$.
This probability is a natural generalization of the persistence probability $R(t)\equiv
R(t,1)$, which corresponds for the walker to always stepping to the right.

For the one dimensional Ising model at zero temperature, $R(t,x)$ was observed to
decay algebraically at large times with an exponent
$\theta(x)$ continuously varying with $x$ \cite{dg}.
For $x=1$, $\theta(1)=\theta$, the usual persistence exponent.
Figure 3 depicts a log-log plot of $R(t,x)$ for
the two dimensional Ising model at zero temperature, $x$ varying from $-1$ to $1$,
while figure 4 depicts the corresponding exponents $\theta(x)$, extracted from figure 
\ref{fig3}.
Let us mention that algebraic decay of $R(t,x)$ was also observed for the diffusion
equation \cite{dg}, and that this quantity and the corresponding exponents
$\theta(x)$ can be exactly computed for the simple model considered in \cite{bbdg}.

\begin{figure}
\begin{center}
\leavevmode
\epsfxsize=\hsize
\epsfbox{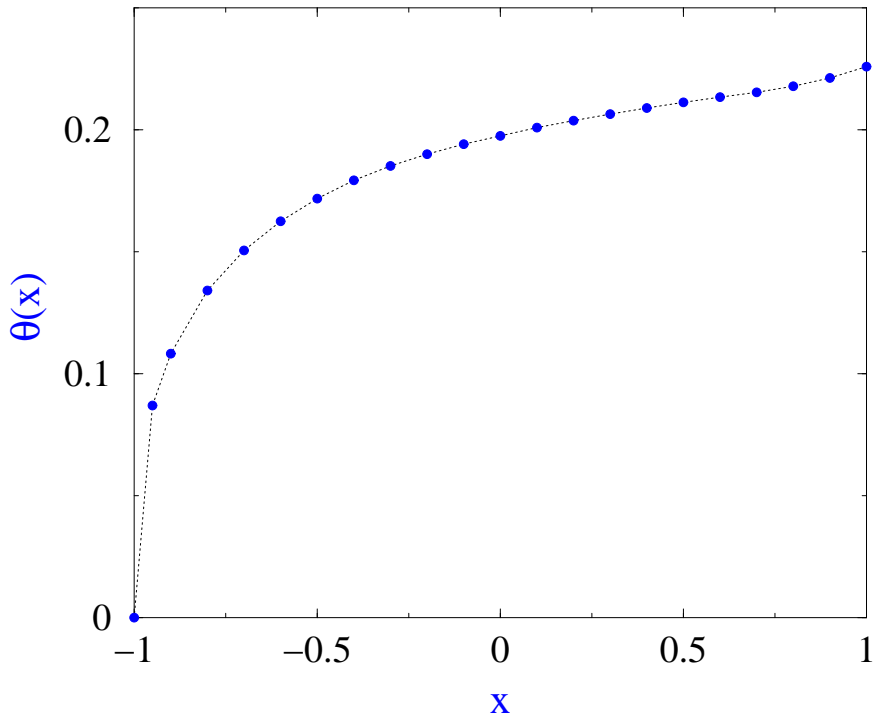}
\end{center}
\caption{
Exponents $\theta(x)$ extracted from figure 3.
\label{fig4}}
\end{figure}

We now address the role of temperature for persistent large deviations.
As is obvious from the first part of this work, if $x>m_0(T)$, then $R(t,x)$ decays
to zero exponentially rapidly. 
On the other hand, for $x<m_0(T)$ one observes algebraic decay of $R(t,x)$, as at
zero temperature. 
Otherwise stated, $x=m_0(T)$ separates two regimes of persistent large deviations,
between exponential and algebraic.
As a consequence, and by analogy with the zero temperature case, one could think of
extracting the persistence exponent at finite temperature from the decay at large times of
$R(t,x)$ when $x\to m_0(T)^-$,
which leads to the formal definition 
\begin{equation}
R(t)=\lim_{x\to m_0(T)^-}R(t,x), \qquad {\text i.e.}\qquad
\theta=\lim_{x\to m_0(T)^-}\theta(x).
\end{equation}
However this definition is not easy to implement in practice.

\begin{figure}
\begin{center}
\leavevmode
\epsfxsize=\hsize
\epsfbox{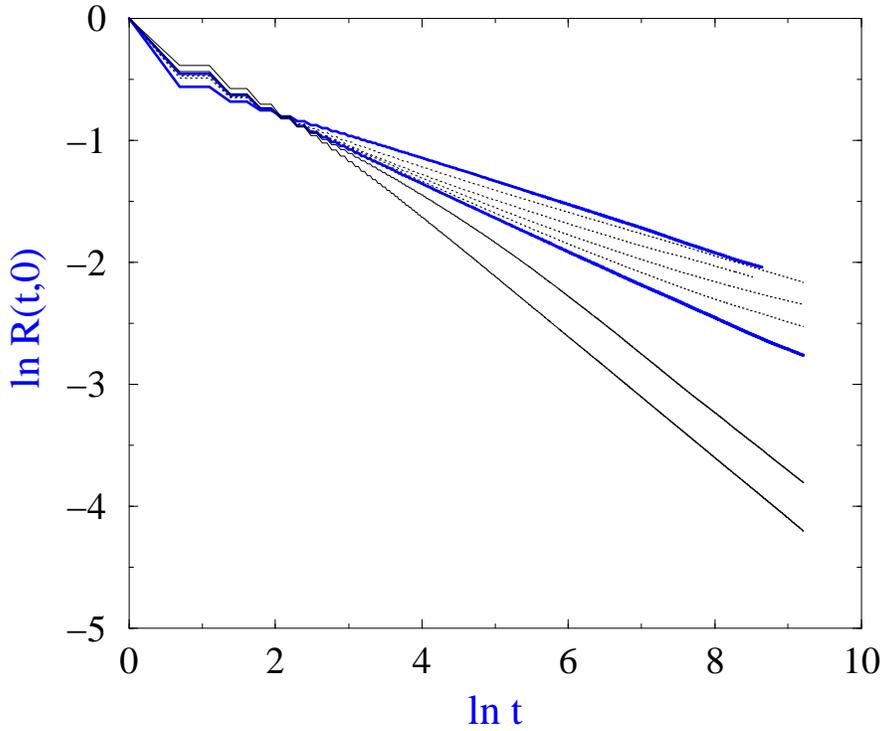}
\end{center}
\caption{
Log-log plot of $R(t,0)$ for temperatures 
$T=0,\,0.8 T_c,\,0.9 T_c,\,$ $0.95 T_c,\,0.98 T_c,\,T_c,\,
1.1 T_c,\,1.5 T_c$, from top to bottom.
(The system size is $1536^2$.)
\label{fig5}}
\end{figure}

We did not investigate the temperature dependence of $\theta(x)$ in all generality.
We restricted our study to the case $x=0$ which
corresponds to
\begin{equation}
R(t,0)=\prob\left(\int_0^t \sigma(t') \d t'\ge0, \forall t'\le t\right) 
=\prob(T_{t'}^{+}\ge T_{t'}^{-},\ \forall t'\le t)
.\label{eqrt0}
\end{equation}
We find that $R(t,0)\sim t^{-\theta(0)}$ with
\begin{equation}
\begin{array}{ll}
\theta(0)={-0.5}\quad &T=\infty,\\
\theta(0)\approx{-0.27}\quad &T=T_c, \\
\theta(0)\approx{-0.20}\quad &T=0. 
\end{array}
\end{equation}
For $T_c<T$, after a crossover, $\theta(0)$ takes the high temperature value $0.5$
while for $T<T_c$ it takes the low temperature value $\approx 0.20$.
(See figure 5.)
The explanation of the value of $\theta(0)$ for $T=\infty$ is simple.
Since spins are independent, identifying  as above $\sigma(t)$, the spin at site
$i$, to the steps of a fictitious one dimensional symmetric random walker,
$R(t,0)$ represents the probability that the walker did not cross the origin up to
time $t$, which is, as is well known, decaying as $t^{-1/2}$
\cite{feller}. 
For decreasing temperatures, spins become more correlated, hence the
exponent
$\theta(0)$ decreases.
Note that  the first passage exponent $\theta(0)$ is defined for $T\ge T_c$, i.e. even
in absence of coarsening. 

This work raises a number of questions.
For instance, what is the temperature dependence of the two time correlation in the
scaling regime, for $T<T_c$?
Is the hypothesis (\ref{eqm2T}) valid?
What is the behaviour of the distribution of $M_t$ at $T_c$ when $t\to\infty$?
At $T_c$, is $\theta(0)$ a new independent critical exponent, or is it related
(equal?) to the persistence exponent $\theta_c$ for the global magnetization
\cite{maj,sirT}? 
Let us mention that for the $3d$ Ising model the quantities
studied here have similar behaviour.
Finally, in our view, an important point of the analysis presented here is that it may
be applied to any coarsening system, since it relies mainly on scaling.

\ack
We wish to thank J-P Bouchaud, I Dornic and J-M Luck for interesting discussions.

\end{document}